\documentclass[aps,prl,twocolumn,nofootinbib,preprintnumbers,amssymb,amsfonts,amsmath,superscriptaddress,showpacs]{revtex4-1}
\usepackage{graphicx}
\usepackage{amsfonts}
\usepackage{amssymb}
\usepackage{amsmath}
\usepackage{hyperref}
\usepackage{slashed}
\usepackage{cancel}
\usepackage{enumerate}
\usepackage{color}
\def\eq#1{{eq.~(\ref{#1})}}

\def\hhref#1{\href{http://arxiv.org/abs/#1}{#1}} 

\definecolor{oucrimsonred}{rgb}{0.6, 0.0, 0.0}
\definecolor{persianblue}{rgb}{0.11, 0.22, 0.73}
\definecolor{forestgreen}{rgb}{0.13,0.35,0.13}
 \hypersetup{colorlinks, citecolor=oucrimsonred, linkcolor=persianblue, urlcolor=oucrimsonred}
\newcommand{\be}{\begin{equation}}
\newcommand{\ee}{\end{equation}}
\newcommand{\bea}{\begin{eqnarray}}
\newcommand{\eea}{\end{eqnarray}}
\newcommand{\nn}{\nonumber}

\begin{document}
\title[]{Natural minimal dark matter}
\date{\today}
\author{Marco Fabbrichesi}
\email{marco@sissa.it}
\affiliation{INFN, Sezione di Trieste, via Valerio 2, I-34137 Trieste, Italy}
\author{Alfredo Urbano}
\email{alfredo.leonardo.urbano@cern.ch}
\affiliation{CERN, Theory division, CH-1211 Gen\`eve 23,
Switzerland}
\begin{abstract}
\noindent  We show how the Higgs boson mass is protected from the potentially  large corrections due to the introduction of minimal  dark matter if the new physics sector is made supersymmetric. The fermionic dark matter candidate (a 5-plet of $SU(2)_L$) is accompanied by a scalar state. The weak gauge sector is made supersymmetric and the Higgs boson is embedded in a supersymmetric multiplet. The remaining standard model states are non-supersymmetric. Non vanishing corrections to the Higgs boson mass only appear at three-loop level and the model is natural for dark matter masses up to 15 TeV---a value larger than the one required by the cosmological relic density. 
The construction presented  stands as an example of a general approach to naturalness that solves  the little hierarchy problem which arises when new physics is added beyond the standard model at an energy scale  around 10 TeV. 

\end{abstract}

\maketitle

\textit{Introduction.} 
Minimal dark matter (MDM)~\cite{Cirelli:2005uq} is  an attractive model  because the stability of the dark matter (DM) candidate is not enforced by an additional {\em ad hoc} symmetry and the coupling to ordinary matter is fixed and equal to that of the weak gauge interaction. The model contains  one or more new particles belonging to  multiplets in a representation of the weak $SU(2)_L$ gauge group that makes their  coupling to standard model (SM) particles by means of renormalizable operators impossible, except for the gauge interaction itself. A fermionic multiplet with $n=5$ is singled out by the simultaneous requirements of containing a stable (neutral) state and preserving the perturbative running of the gauge coupling---that would be destroy by too large a representation. 

This model  suffers of a (mild) problem of naturalness~\cite{strumia} insofar as the mass of the DM candidate must be around 10 TeV to satisfy the current relic abundance constraint~\cite{Cirelli:2007xd}. Such a value turns into a correction to the Higgs boson mass roughly one order of magnitude larger than its value and therefore give rise to a (little) hierarchy problem.

In this work we show how MDM can be made natural by making the DM sector and its interaction with the SM supersymmetric while leaving the SM itself non supersymmetric.

\vskip 1.3em
\textit{Naturalness.}
Naturalness~\cite{nat} requires quantum corrections to the Higgs boson mass to be of the same order of the mass itself. It is not a physical principle as long as the Higgs boson mass is an input in the model and not a computable parameter.  It is just a requirement we add on a model to satisfy a prejudice we entertain about the size of radiative corrections. The naturalness requirement cannot be stated in  general  because it depends on the specific kind of new physics one has introduced. It is best expressed in terms of finite corrections without reference to cutoff dependent quantities that render the issue moot. In the absence of new physics the SM by itself is natural~\cite{natural,strumia}. When new physics is added at an energy  threshold  significantly larger than the electroweak (EW) scale, corrections proportional to such a scale and the coupling of the new states to the Higgs boson mass appear, the size of which require an unnatural cancellation in the definition of the mass parameter.  

For very large thresholds a serious problem of hierarchy is present. As shown by the example of GUT~\cite{natgut}, this problem can be solved by making the theory supersymmetric.
For more modest energy scales, a little hierarchy problem might be identified and solved by a partial implementation of supersymmetry (SUSY) that only includes the new physics sector and the Higgs boson. This approach was recently emphasised  in~\cite{Fabbrichesi:2015zna} and it is here applied to the MDM hierarchy problem (for a similar approach to the problem of  naturalness and DM see also \cite{Heikinheimo:2013xua}).
 
The correction to the Higgs boson mass in the MDM model comes at the two-loop level because  MDM only interacts with the SM particles via the $SU(2)_L$ gauge bosons. It is given by
 \bea
\left.  \delta m^2_h \right|_{\psi} &=& 30\,  \frac{ g^4 M^2}{(4 \pi)^4} \left[ 6 \ln \frac{M^2}{\mu^2} - 1 \right] ,  \\
\left.  \delta m^2_h  \right|_{\phi } &=& 
 - 30\,  \frac{ g^4 M^2}{(4 \pi)^4} \left[ \frac{3}{2}  \ln^2 \frac{M^2}{\mu^2} + 2 \ln \frac{M^2}{\mu^2}- \frac{7}{2} \right] , \label{2loop} \nn
 \eea
respectively, for a Majorana $\psi$ and a scalar  $\phi$  DM candidate, both with $n=5$ and hypercharge $Y=0$. Notice that the two contributions do not cancel against each other, not even in their non-polynomial parts.
Naturalness would require that this correction be of the same order of the Higgs boson mass $m_h$. By taking the matching scale  at its natural value $\mu = m_h$, we find
that the largest value of $M$ satisfying this requirement is 1.5 TeV in the fermionic case and 4.2 TeV in the scalar case. In ref.~\cite{strumia} lower values for this limiting masses are found, namely $M\leq 400$ GeV because the matching scale is taken to be the Planck mass. Such a choice does not seem justified in as much as the only threshold assumed in the model is the new states of mass $M$. 

We  thus have a little hierarchy problem because the DM candidate must have $M=10$ TeV~\cite{Cirelli:2007xd} in order to 
saturate the cosmological relic density. Such a problem, given the numbers involved, can  be waved away by  a (rather mild)  fine tuning or by increasing the number of DM candidates. On the other hand, one can use the naturalness requirement in an heuristic way to define an improved  model.

\vskip 1.3em
\textit{The model.}
We embed the MDM $n=5$ states  and the SM Higgs boson into three chiral supermultiplets
\bea
\Phi_{\rm DM}^{A} &=& \phi^{A} + \theta\cdot \psi^{A} + F^{A} \theta^2  \, ,\\
\Phi_{H_{u,d}}^{a} &=& H_{u,d}^a + \theta\cdot \tilde{h}_{u,d}^a + G_{u,d}^a \theta^2\, ,
\eea
where $A$ and $a$ are $SU(2)_L$ indices: $\Phi_{\rm DM}^{\rm A}$ belongs to the 5-dimensional (real) representation of $SU(2)_L$ with zero hypercharge
while the two Higgs doublets $\Phi_{H_{u,d}}^{a} $ belong to the fundamental of $SU(2)_L$ with hypercharges $Y_u = -1/2$, $Y_d = 1/2$. 
$\psi$ and $\phi$ are, respectively, the fermionic and scalar DM candidate.
By construction, $\psi$ is a two-component left Weyl fermion, while $\phi$ is a complex scalar.
We adopt the following parametrization (before  EW symmetry breaking) for the four neutral and two charged degrees of freedom:
 \begin{eqnarray}
 H_u &=&
\left(
\begin{array}{c}
\frac{1}{\sqrt{2}} (H^0 c_{\alpha} - h_0  s_{\alpha} +i A^0 s_{\beta} -iG^0 c_{\beta})  \\
  H^- s_{\beta} - G^- c_{\beta} 
\end{array}
\right)\,,\label{eq:Hup}\nonumber\\
 H_d &=&
\left(
\begin{array}{c}
 H^+ c_{\beta} + G^+ s_{\beta}  \\
\frac{1}{\sqrt{2}} (H^0 s_{\alpha} + h_0 c_{\alpha} +i A^0 c_{\beta} + iG^0 s_{\beta})
\end{array}
\right)\,,\label{eq:Hdown}
 \end{eqnarray}
 where $s_{x} \equiv \sin x$, $c_{x} \equiv \cos x$, $t_{x} \equiv \tan x$.
 In eq.~(\ref{eq:Hdown}) $\alpha$, $\beta$ are two mixing angles and the neutral $h_0$ component is the physical Higgs of the SM.
 Finally, $F^A$ and $G^a_{u,d}$ are 
 non-dynamical complex auxiliary fields, and they feature the same $SU(2)_L$ transformation properties
 of the chiral superfield they belong to.

 The supersymmetric lagrangian is that of the gauged Wess-Zumino model~\cite{Wess:1973kz}
\bea
\mathcal{L}_{WZ} & = &  \int d^2 \theta d^2 \bar \theta \left[  \Phi_{\rm DM}^\dag \, e^{2gV}\, \Phi_{\rm DM} 
  + \Phi_{H_{k}}^\dag \, e^{2gV}\,  \Phi_{H_{k}} \right]  \label{WZ} \\  
  & + &
\left[ \frac{1}{2}\int d^2 \theta \: {\rm Tr}(\textsc{W}\cdot \textsc{W}) +  \int d^2 \theta \: {\cal W}(\Phi_{\rm DM}) + h.c. \right]\,,\nn
\eea
and it consists of three parts. The first term in $\mathcal{L}_{WZ}$, in which the sum over $k=u,d$ is implicit, is the usual non-abelian K\"ahler potential. The vector superfield is defined in the adjoint representation of $SU(2)_L$, $V_{AB} = V^{\alpha}(T^{\alpha}_{\textbf{3}})_{AB}$.
In the cartesian basis the generators of the adjoint representation of $SU(2)_L$ are explicitly given by $(T^{\alpha}_{\textbf{3}})_{AB} = -i\epsilon_{\alpha AB}$.
In the Wess-Zumino gauge the vector superfield is
\be
\label{eq:VectorSF}
V^{\alpha} = \theta\cdot (\sigma^{\mu}\bar{\theta})\,W^{\alpha}_{\mu} + \bar{\theta}^2\,\theta\cdot \lambda^{\alpha} + \theta^2\,\bar{\theta}\cdot \bar{\lambda}^{\alpha}
+\frac{1}{2}\theta^2\,\bar{\theta}^2 D^{\alpha}~,
\ee
where $W_\mu^{\alpha}$ and $\lambda^{\alpha}$ are, respectively, the $SU(2)_L$ gauge bosons and the corresponding gaugino triplet.
Finally, $D^{\alpha}$ is a non-dynamical real auxiliary field.
The supersymmetric field strength is $\textsc{W}_{s} = -(1/8g)\bar{D}^2(e^{-2gV}D_{s}e^{2gV})$, where $s$ is a spinorial index and 
the supersymmetric covariant derivative 
is $D_{s} = \partial_{s} - i(\sigma^{\mu})_{s\dot{s}}\bar{\theta}^{\dot{s}}\partial_{\mu}$. 

The superpotential in \eq{WZ} does not contain interactions and is given by
\begin{equation}\label{pot}
{\cal W}(\Phi_{\rm DM}) =   \frac{M}{2}(\epsilon_{\textbf{5}})_{AB} \Phi_{\rm DM}^{A}\Phi_{\rm DM}^{B}\,, 
\end{equation}
where the  parameter $M$ is the mass of the DM candidate. The isospin invariant coupling 
is guaranteed in eq.~(\ref{pot}) by the presence of the symmetric tensor $\epsilon_{\textbf{5}}$ realizing the equivalence between the 
5-dimensional representation of $SU(2)_L$ and its conjugate. Considering  the representation $\textbf{n}$ of $SU(2)_L$ with generators $T^{a=1,2,3}_{\textbf{n}}$,
the tensor $\epsilon_{\textbf{n}}$ is defined by the equivalence relation $\epsilon_{\textbf{n}} T^{a}_{\textbf{n}} (\epsilon_{\textbf{n}})^{-1} = -(T^{a}_{\textbf{n}})^*$. 
If $\textbf{n}$ is even, $\epsilon_{\textbf{n}}$ is antisymmetric; in this case (pseudo-real representation
 of $SU(2)_L$) the mass term in eq.~(\ref{pot}) vanishes.

In components, the supersymmetric lagrangian of the model takes the form
\begin{widetext}
\begin{align}
\label{eq:FullLagrangian}
\mathcal{L}_{WZ} & = (D_{\mu}\phi)^{\dag}(D^{\mu}\phi) - M^2|\phi|^2 
-\frac{g^2}{2}(\phi^{\dag}T^A_{\textbf{5}}\phi)(\phi^{\dag}T^A_{\textbf{5}}\phi)
+ \bar{\psi}i\bar{\sigma}^{\mu}(D_{\mu}\psi) -\frac{M}{2} 
(\epsilon_{\textbf{5}})_{AB}\left(\psi^A\cdot \psi^B + \bar{\psi}^A\cdot \bar{\psi}^B \right) \nonumber \\
& + (D_{\mu}H_k)^{\dag}(D^{\mu}H_k) 
-\frac{g^2}{2}
(H_k^{\dag} T^a_{\textbf{2}} H_k)(H_j^{\dag} T^a_{\textbf{2}} H_j)
+ \bar{\tilde{h}}_k i\bar{\sigma}^{\mu}(D_{\mu}\tilde{h}_k) - \frac{1}{4}W_{\mu\nu}^{\alpha}W^{\alpha,\mu\nu} + 
\bar{\lambda}i\bar{\sigma}^{\mu}(D_{\mu}\lambda) \nonumber \\
& -\sqrt{2}g\left[ 
(\phi^{\dag} T^{\alpha}_{\textbf{5}} \psi)\cdot \lambda^{\alpha} + \bar{\lambda}^{\alpha} \cdot (\bar{\psi} T^{\alpha}_{\textbf{5}} \phi)
\right] -\sqrt{2}g\left[ 
(H_k^{\dag} T^{\alpha}_{\textbf{2}} \tilde{h}_k)\cdot \lambda^{\alpha} + \bar{\lambda}^{\alpha} \cdot (\bar{\tilde{h}}_k T^{\alpha}_{\textbf{2}}  H_k)
\right]
-g^2(H_k^{\dag} T^a_{\textbf{2}} H_k)(\phi^{\dag}T^a_{\textbf{5}}\phi)~,
\end{align}
\end{widetext}
where the ordinary $SU(2)_L$ covariant derivatives 
are 
$D_{\mu}\phi = \partial_{\mu}\phi + igW_{\mu}^{\alpha}(T^{\alpha}_{\textbf{5}} \phi)$,  $D_{\mu}\psi = \partial_{\mu}\psi + igW_{\mu}^{\alpha}(T^{\alpha}_{\textbf{5}} \psi)$, 
$D_{\mu}H_k = \partial_{\mu}H_k + igW_{\mu}^{\alpha}(T^{\alpha}_{\textbf{2}} H_k)$, and 
$D_{\mu}\lambda = \partial_{\mu}\lambda + igW_{\mu}^{\alpha}(T^{\alpha}_{\textbf{3}} \lambda)$.
For the fundamental $SU(2)_L$ representation we have $T^{\alpha}_{\textbf{2}} = \sigma^{\alpha}/2$, with $\sigma^{\alpha = 1,2,3}$ the
usual Pauli matrices.
Thanks to the supersymmetric structure, the model---despite the introduction of new fields---preserves its simplicity:
all the coupling are set by gauge interactions, and the only free parameter is the mass of the chiral supermultiplet $\Phi_{\rm DM}$.
The fermionic mass term in the first line of eq.~(\ref{eq:FullLagrangian}) is not diagonal.
It can be easily diagonalized as follows
\begin{equation}
 -\frac{M}{2}
\left[
(\psi_L^{A})^T\mathcal{C}(\epsilon_{\textbf{5}})_{AB}\psi_L^B + h.c.
\right] = -\frac{M}{2}\overline{\Psi^{A}}\Psi^A~,
\end{equation}
where on the l.h.s.
we just rewrote the mass term in four-component notation while the Majorana mass eigenstates on the r.h.s.
are defined by $\Psi \equiv U^{\dag}\psi_L + (U^{\dag}\psi_L)^{C}$. The unitary transformation matrix is
implicitly defined via $U^T \epsilon_{\textbf{5}} U = \textbf{1}$, 
 and the charge conjugation is $\psi^C = \mathcal{C}(\overline{\psi})^T$. 
 
\begin{figure}[!t!]
\centering
 \includegraphics[width = 0.45 \textwidth]{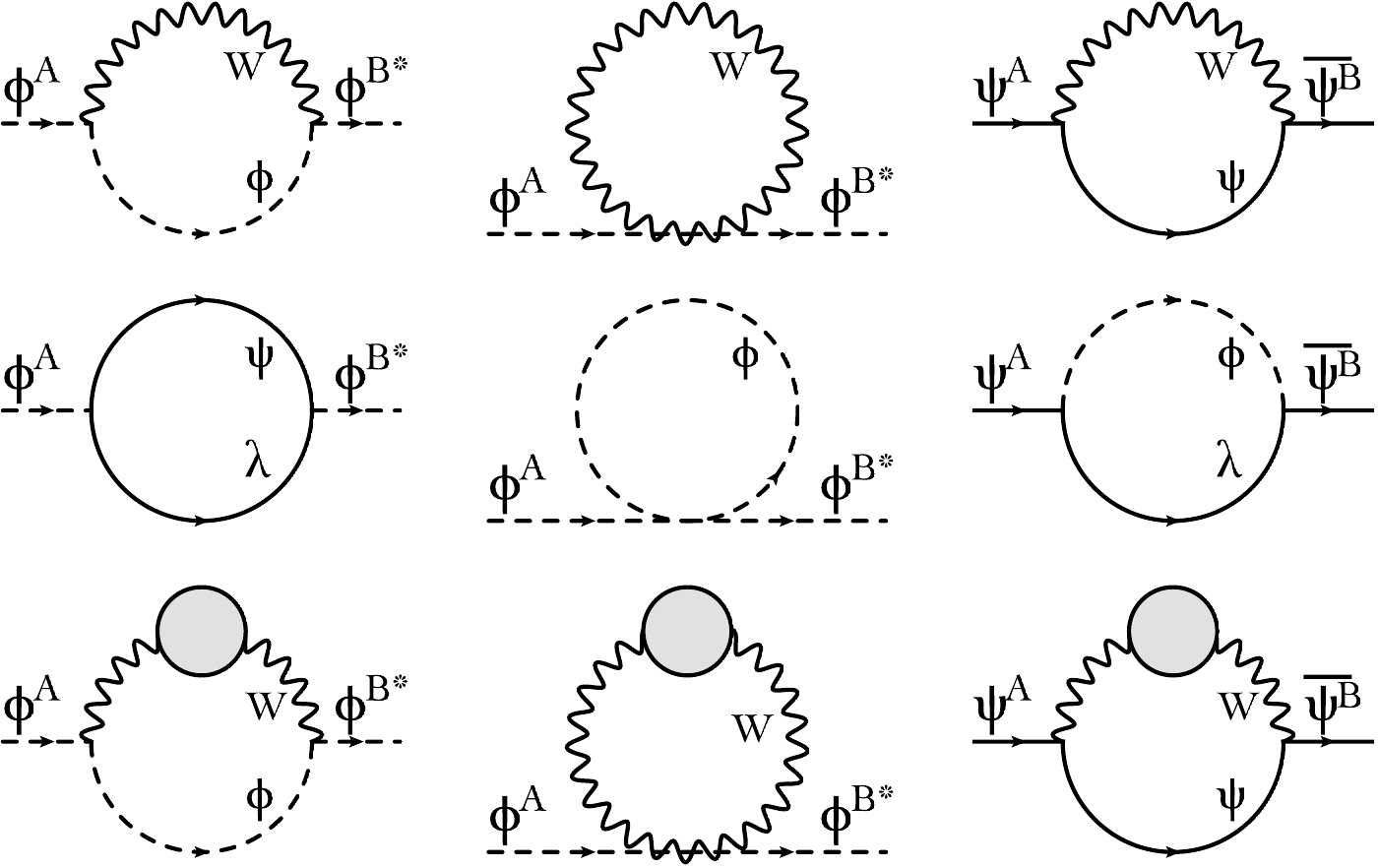}
\caption{ \textit{
First two rows: One-loop SUSY-preserving diagrams involved in the renormalization of the mass $M$.
Lower row: Two-loop diagrams responsible---via SUSY-breaking SM interactions---for the mass splitting between the scalar and fermion multiplet;
the gray blob represents the contribution of non-supersymmetric SM particles.
}}
\label{fig:SUSYsplit}
\end{figure}

The complete lagrangian for the model is given by
\begin{align}\label{eq:FullLagrangian2}
\mathcal{L}_{\rm nMDM}  & = 
{\cal L}_{SM} + {\cal L}_{WZ} - \frac{1}{2} \tilde m_\lambda (\lambda^{\alpha}\cdot \lambda^{\alpha} + h.c.)  \nonumber \\
 & + \mu (\epsilon_{\alpha\beta}\tilde{h}_{u}^{\alpha}\cdot \tilde{h}_d^{\beta} + h.c.)~, 
\end{align}
where ${\cal L}_{SM}$ is the (non-supersymmetric) SM lagrangian, and the last term in the first line gives mass to the gauginos. 
The two higgsinos are coupled via a $\mu$-term, with $\epsilon \equiv i\sigma^2$.

\vskip 1.3em
\textit{Physical states and their masses.}
The two mass parameters $\tilde m_\lambda$ and $\mu$ (taken to be real for simplicity) are  of the order of the EW scale and do not therefore give rise to unnatural corrections to the Higgs boson mass. 
By defining $\tilde{W}^0 \equiv \lambda^3$, $\tilde{W}^{\pm}\equiv (\lambda^1 \mp i\lambda^2)/\sqrt{2}$, and 
introducing the analogous of the neutralino
$\tilde{G}^0 \equiv (\tilde{W}^0, \tilde{h}_d^0, \tilde{h}_u^0)^T$ and chargino
 $\tilde{g}^+ \equiv (\tilde{W}^+, \tilde{h}_d^+)^T$, $\tilde{g}^- \equiv (\tilde{W}^-, \tilde{h}_u^-)^T$ states,
   we extract from eq.~(\ref{eq:FullLagrangian2}) the following mass terms
\begin{eqnarray}
\mathcal{L}_{\chi^0} &=& -\frac{1}{2}(\tilde{G}^0)^T
\left(
\begin{array}{ccc}
\tilde{m}_\lambda  & -\frac{gv}{2}s_{\beta}  & \frac{gv}{2} c_{\beta}  \\
-\frac{gv}{2} s_{\beta}&  0 & -\mu  \nonumber \\
 \frac{gv}{2} c_{\beta}  &  -\mu &  0 
\end{array}
\right)\tilde{G}^0 + h.c.\\
& \equiv &  -\frac{1}{2}(\tilde{G}^0)^T \mathcal{M}_{\chi^0}\tilde{G}^0 + h.c.\,,\\
\mathcal{L}_{\chi^{\pm}} &=& -\frac{1}{2}\left[
(\tilde{g}^+)^T
\mathcal{M}^T
\tilde{g}^- +
(\tilde{g}^-)^T
\mathcal{M}
\tilde{g}^+ 
\right] + h.c.~,
\end{eqnarray}
with   $\mathcal{M} = \left(
\begin{array}{cc}
\tilde{m}_\lambda  & gv s_{\beta}/\sqrt{2}   \\
gv c_{\beta}/\sqrt{2}  &  \mu    
\end{array}
\right)$. The off-diagonal entry in $\mathcal{M}$ is due, after EW symmetry breaking with $\langle H_d\rangle = (0,v s_{\beta}/\sqrt{2})^T$, $\langle H_u\rangle = (v c_{\beta}/\sqrt{2}, 0)^T$, to the supersymmetric Yukawa interactions in eq.~(\ref{eq:FullLagrangian}).
The neutralino mass matrix can be diagonalized via the unitary transformation $N^T \mathcal{M}_{\chi^0} N = {\rm diag}(m_1, m_2, m_3)$, and we denote
the corresponding mass eigenstates as $\tilde{\chi}^0 = N^{\dag}\tilde{G}^0$.
Since $\mathcal{M}^T \neq \mathcal{M}$ (unless $t_{\beta}=1$) two distinct unitary transformations, $V$ and $W$, are needed for its diagonalization. 
We denote the corresponding mass eigenstates as $\tilde{\chi}^{+} = V^{\dag} \tilde{g}^+$, $\tilde{\chi}^{-} = W^{\dag} \tilde{g}^-$, with 
$\tilde{\chi}^{\pm} = (\tilde{\chi}_1^{\pm}, \tilde{\chi}_2^{\pm})^T$. 
  
 At the tree level, and before the EW symmetry breaking, the scalar and fermion multiplets have the same mass $M$.
 At one loop, SUSY preserves the degeneracy between the multiplets, as a consequence of the non-renormalization theorem~\cite{Iliopoulos:1974zv}.
 More explicitly, it is possible to show that the mass renormalization  induced
 by the one-loop diagram in the first two rows of fig.~\ref{fig:SUSYsplit} is exactly the same for the scalar and fermion multiplet.
 The (non-supersymmetric) SM lagrangian introduces an explicit breaking of SUSY. 
 All the cancellations and properties inherent to the supersymmetric structure of the model fail when higher order corrections involved non-supersymmetric SM particles are considered.
 At two loops, the mass degeneracy between $\phi^A$ and $\psi^A$ is broken, and 
we show  in the lower row of fig.~\ref{fig:SUSYsplit}  the typical diagrams responsible for this effect.
 
The size of this two-loop correction is 
 \begin{equation}\label{eq:TwoLoopSplitting}
 \Delta M_{\cancel{{\rm SUSY}}}^{(\phi,\psi)} \simeq 
 \frac{g^2}{(16\pi^2)} M
   \simeq 2.5 \times 10^{-3} \; M\, .
\end{equation}
The actual value and sign of this correction depend on the details of the two-loop computation, and 
 in principle it can be different for the scalar and fermion multiplet.
The mass shift   $\Delta M_{\cancel{{\rm SUSY}}}^{(\phi,\psi)}$ is of the entire
multiplet---weather scalar or fermion---and it does not introduce any split between the single components. 

A mass splitting within the components of the multiplet is introduced {\it i)} at one-loop 
by SM EW interactions~\cite{Cirelli:2005uq},  {\it ii)} at one-loop by supersymmetric interactions, and {\it iii)} at the tree level, after EW symmetry breaking, by 
the presence of the operators $(H_{u,d}^{\dag} T^a_{\textbf{2}} H_{u,d})(\phi^{\dag}T^a_{\textbf{5}}\phi)$ in eq.~(\ref{eq:FullLagrangian})~\cite{Cirelli:2005uq,Hambye:2009pw}.
The first correction---generated by the diagrams in the first row in fig.~\ref{fig:SUSYsplit}---does not depend on the spin, and it splits the components of $\phi^A$ and $\psi^A$ in the same way; 
the third correction, on the other hand, only affects the scalar multiplet. We have
 \begin{eqnarray} \label{split}
 M_{\phi}^{(Q)} &=& M_{\phi} + Q^2 \Delta M_g + \Delta M^{\phi}_{\rm SUSY} - Q(s_{\beta}^2 - c_{\beta}^2)\frac{g^2 v^2}{8 M_{\phi}}\, ,\nn \\
  M_{\psi}^{(Q)} &=& M_{\psi} + Q^2 \Delta M_g  + \Delta M^{\psi}_{\rm SUSY}~,
 \end{eqnarray}
 where the electric charge $Q  = \pm 2, \pm 1, 0$ distinguishes  the components of the multiplets, and 
 \be
 \Delta M_g \simeq 166\; \mbox{MeV}\, ,
 \ee
  is the splitting induced by the EW interactions; finally, as discussed before, $M_{\phi}  = M + \Delta M_{\cancel{{\rm SUSY}}}^{(\phi)}$, $M_{\psi}  = M + \Delta M_{\cancel{{\rm SUSY}}}^{(\psi)}$. 
  The mass splitting in eq.~(\ref{split}) induced by supersymmetric interactions---the diagrams in the second row in fig.~\ref{fig:SUSYsplit}---is given by
  \begin{align}\label{eq:SUSYsplit}
\Delta M_{\rm SUSY}^{\psi} = & \frac{g^2 M}{16\pi^2}\left[
\frac{1}{2}(Q^2 - Q)V_{1k}V^{\dag}_{k1}R_{\tilde{\chi}^+_k}^2(1 + \ln R_{\tilde{\chi}^+_k}^2)\right.  \nn \\ & \left. +
\frac{1}{2}(Q^2 + Q)W_{1k}W^{\dag}_{k1}R_{\tilde{\chi}^-_k}^2(1 + \ln R_{\tilde{\chi}^-_k}^2)\right.  \nn \\ & \left. 
 - Q^2(N_{1k}N_{k1}^{\dag})R_{k}^2(1 + \ln R_{k}^2) 
\right]\,,\\
\Delta M_{\rm SUSY}^{\phi} = & -\frac{g^2 M}{16\pi^2}\left[
\frac{1}{2}(Q^2 - Q)V_{1k}V^{\dag}_{k1}R_{\tilde{\chi}^+_k}^2(3 - \ln R_{\tilde{\chi}^+_k}^2)\right.  \nn \\ & \left. +
\frac{1}{2}(Q^2 + Q)W_{1k}W^{\dag}_{k1}R_{\tilde{\chi}^-_k}^2(3 - \ln R_{\tilde{\chi}^-_k}^2) \right.  \nn \\ & \left. - Q^2(N_{1k}N_{k1}^{\dag})R_{k}^2(3 - \ln R_{k}^2) 
\right]~,\label{eq:SUSYsplit2}
\end{align}
where $R_a \equiv m_a/M$. In the square brackets, the first two terms (last term) are (is)
generated by loop exchange of charginos (neutralinos). Notice that in the absence of EW symmetry breaking 
there is no mixing in the mass matrices $\mathcal{M}_{\chi^0}$ and $\mathcal{M}$; as a consequence, the neutral and charged contributions in 
eqs.~(\ref{eq:SUSYsplit},\ref{eq:SUSYsplit2}) cancel each other out. An analogous  cancellation is valid also in the pure EW sector, and $ \Delta M_g =0$ in the unbroken EW phase. 
 On the other hand, after EW symmetry breaking,  $\Delta M_{\rm SUSY}^{\phi,\psi} = 0$ vanishes in the limit $t_{\beta} = 1$ since neutral and charged degrees of freedom 
 are diagonalized in the same way.
 
 The splitting $Q^2 \Delta M_g$ makes all the charged components heavier than the neutral one; the mass splitting induced by the Higgs vacuum expectation value $v$, 
 on the contrary, 
 depends on the sign of the difference $s_{\beta}^2 - c_{\beta}^2$, and for $s_{\beta}^2 - c_{\beta}^2 > 0$
 makes the positively (negatively) charged components lighter (heavier) than the neutral one. The coupling in front of
 the operators $(H_{u,d}^{\dag} T^a_{\textbf{2}} H_{u,d})(\phi^{\dag}T^a_{\textbf{5}}\phi)$
  is fixed by SUSY to be equal to $g^2$, and cannot be 
 neglected as usually assumed in MDM-inspired scalar models. 
 Furthermore, the correction induced by $(H_{u,d}^{\dag} T^a_{\textbf{2}} H_{u,d})(\phi^{\dag}T^a_{\textbf{5}}\phi)$ dominates if compared to $\Delta M_{\rm SUSY}^{\phi,\psi}$ since the latter 
 has an extra suppression of order $gv/M$.
 
 As far as the scalar multiplet is concerned, therefore, 
 in order to avoid the presence of a charged particle as the lightest component of the multiplet 
it is necessary that $M_{\phi} > (s_{\beta}^2 - c_{\beta}^2)g^2v^2/8\Delta M_g$. In turn, this condition can be recast into
a constraint on $\beta$ for a given $M_{\phi}$. 

The model with only one chiral multiplet $\Phi^a_{H_d}$ (with the Higgs boson as its scalar component), while more attractive because simpler, receives a tree-level EW correction that is not suppressed by the mixing we have in the presence of a second scalar doublet and therefore may produce a  scalar DM candidate with a charged component lighter than the neutral one.

Finally, we remark 
that all the splittings generated by the Higgs  vacuum expectation value---hence either $\Delta M_g$, $\Delta M_{\rm SUSY}^{\phi,\psi}$ or the splitting induced by  $(H_{u,d}^{\dag} T^a_{\textbf{2}} H_{u,d})(\phi^{\dag}T^a_{\textbf{5}}\phi)$---depend on the temperature, since they vanishes for $T > T_{\rm c}$---with 
$T_{\rm c}$ the critical temperature of the EW phase transition---when the $SU(2)_L$ symmetry is restored.
We shall return  to this point when we  compute  the relic density.

\vskip 1.3em
\textit{Stability of the DM candidates.}
Non-renormalizable operators could in principle open new decay channels and make MDM unstable. In the framework we are following, this is not possible without introducing a new scale and therefore negate the entire approach. 

By assuming a unique threshold scale, we know the UV completion of the the SM and therefore know what non-renormalizable operators can or cannot be present. In our case, there is no way to construct operators leading to  DM decay like
\be
\frac{1}{M} \phi  H H H^* H^* \quad \mbox{or} \quad \frac{1}{M^2} \psi L H H^*
\ee
by means of the lagrangian in \eq{eq:FullLagrangian2} and both $\phi$ and $\psi$ are stable. Therefore, as opposed to the original MDM model, 
both multiplets  may contain a  DM  candidate (the neutral component of $\psi^A$ and $\phi^A$), and  it is their combined abundance that has to be compared to the cosmological bound.

As discussed in the previous section, interactions mediated by non-supersymmetric SM particles break the degeneracy between the multiplets $\psi^A$ and $\phi^A$.
As a consequence, the trilinear Yukawa interaction $\sqrt{2}g\left[ 
(\phi^{\dag} T^{\alpha}_{\textbf{5}} \psi)\cdot \lambda^{\alpha} + h.c.\right]$ in eq.~(\ref{eq:FullLagrangian}) could lead to decays $\psi \to \phi\lambda$ (or $\phi\to \psi\lambda$, depending on which one
between the two multiplets
 turns out to be the lightest once the corrections in eq.~(\ref{eq:TwoLoopSplitting}) are properly computed) with final state gauginos. 
 However, the size of the typical mass splitting in eq.~(\ref{eq:TwoLoopSplitting}) is far too small to kinematically open these decay channels since the mass of the gaugino in the final state is of the order of the EW scale.

\begin{figure}[!t!]
\centering
 \includegraphics[width = 0.45 \textwidth]{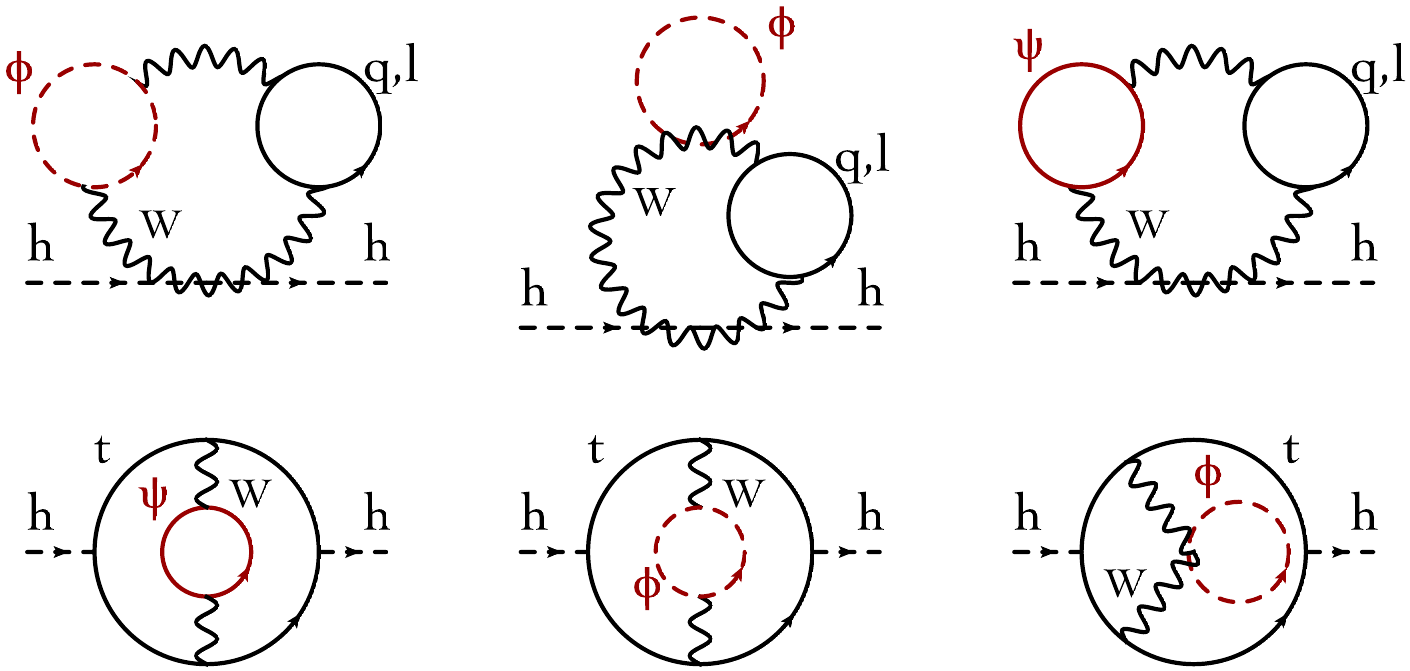}
\caption{ \textit{Representative diagrams giving rise to three-loop order corrections to the Higgs boson mass proportional to $M^2$.
}}
\label{fig:HiggsCorrections}
\end{figure}

\vskip 1.3em
\textit{Corrections to the Higgs boson mass.} Because of the supersymmetric structure, the first correction to the Higgs boson mass comes at three-loop level (see fig.~\ref{fig:HiggsCorrections}) as opposed to the two-loop result in \eq{2loop}. 
At two loops, the cancellation of large quadratic corrections proportional to $M^2$
 is guaranteed by the supersymmetric structure of the theory.

\begin{figure}[!t!]
\centering
 \includegraphics[width = 0.45 \textwidth]{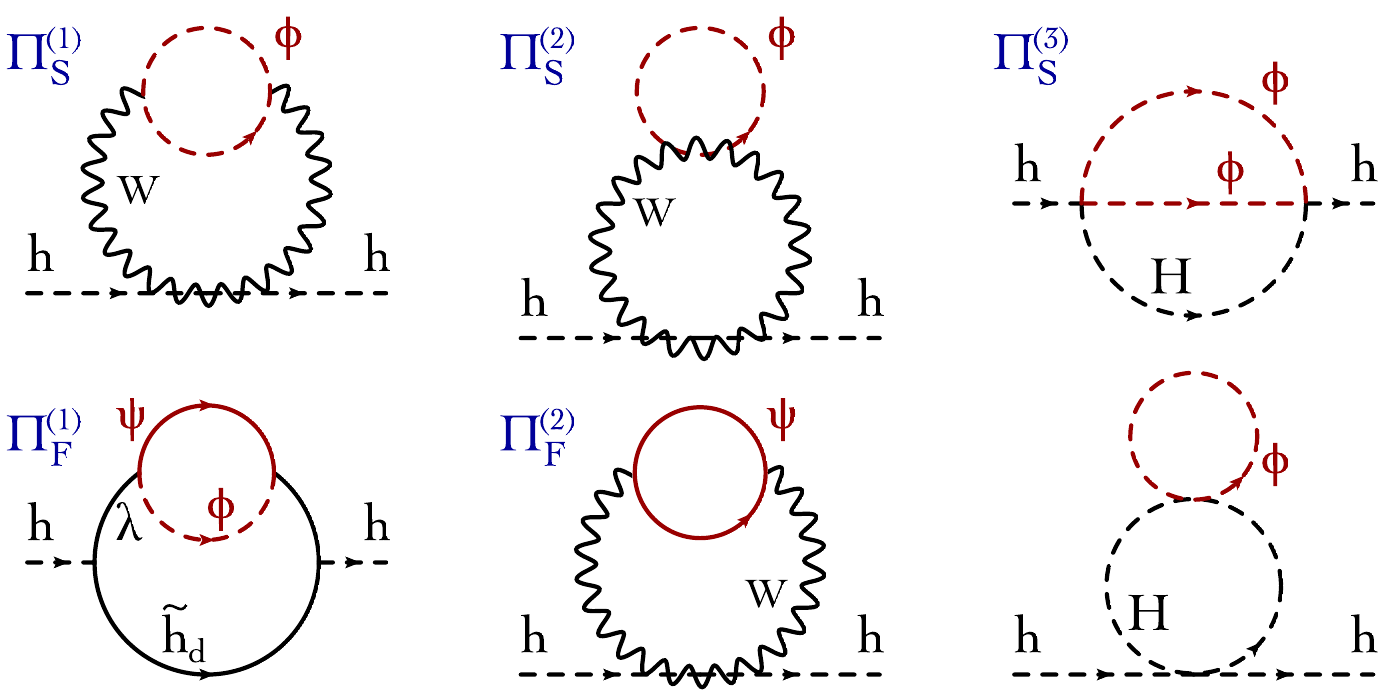}
\caption{ \textit{
Two-loop corrections to the Higgs boson self-energy  
involved in the cancellation of the corrections proportional to $M^2$.
For each diagram we mark in red the contribution of the MDM chiral multiplet.
}}
\label{fig:TwoLCanc}
\end{figure}

For simplicity, we illustrate the cancellation mechanism in the limit $\alpha = 0$, $\beta = \pi/2$. 
In this limit
there is no mixing between the components of the two Higgs doublets, and the SM Higgs can be identified with the real part of the neutral component of $H_d$. 
The two-loop diagrams involved in the cancellation are shown in fig.~\ref{fig:TwoLCanc}. 
Since we are only interested in the renormalization of the Higgs boson mass induced by the MDM scale $M$, 
we compute the two-point function of the Higgs boson setting the external momentum to zero, 
and we work in the massless EW theory. 
Moreover we work in the Landau gauge, where diagrams with gauge bosons attached to external scalar lines with zero momentum vanish.
We find the following contributions (see fig.~\ref{fig:TwoLCanc} for the corresponding notation):
\begin{eqnarray}\label{eq:TwoLoopCanc}
i\Pi_{S}^{(1)} &=& -\frac{ig^4\mathcal{C}_{\textbf{5}}}{4}
\int
\frac{d^4k_1}{(2\pi)^4}
\frac{d^4k_2}{(2\pi)^4}
\frac{4k_2^2 - 4(k_1\cdot k_2)/k_1^2}{
\mathcal{D}_{k_1}^{\{2,0\}}
\mathcal{D}_{k_2}^{\{1,M\}}
\mathcal{D}_{k_2 - k_1}^{\{1,M\}}
}~,\nonumber \\
i\Pi_{S}^{(2)} &=& \frac{ig^4\mathcal{C}_{\textbf{5}}}{4}
\int
\frac{d^4k_1}{(2\pi)^4}
\frac{d^4k_2}{(2\pi)^4}
\frac{6}{
\mathcal{D}_{k_1}^{\{2,0\}}
\mathcal{D}_{k_2}^{\{1,M\}}
}~,\nonumber \\
i\Pi_{S}^{(3)} &=& \frac{ig^4\mathcal{C}_{\textbf{5}}}{4}
\int
\frac{d^4k_1}{(2\pi)^4}
\frac{d^4k_2}{(2\pi)^4}
\frac{1}{
\mathcal{D}_{k_1}^{\{1,0\}}
\mathcal{D}_{k_2}^{\{1,M\}}
\mathcal{D}_{k_2-k_1}^{\{1,M\}}
}~,\nonumber \\
i\Pi_{F}^{(1)} &=& -\frac{ig^4\mathcal{C}_{\textbf{5}}}{4}
\int
\frac{d^4k_1}{(2\pi)^4}
\frac{d^4k_2}{(2\pi)^4}
\frac{8k_1\cdot k_2}{
\mathcal{D}_{k_1}^{\{2,0\}}
\mathcal{D}_{k_2}^{\{1,M\}}
\mathcal{D}_{k_2-k_1}^{\{1,M\}}
}~,\nonumber \\
i\Pi_{F}^{(2)} &=& \frac{ig^4\mathcal{C}_{\textbf{5}}}{4}
\int
\frac{d^4k_1}{(2\pi)^4}
\frac{d^4k_2}{(2\pi)^4}\times \nonumber \\
&&
\frac{
- 4(k_1\cdot k_2)/k_1^2 + 6k_1\cdot k_2 -2k_2^2 +6M^2
 }{
\mathcal{D}_{k_1}^{\{2,0\}}
\mathcal{D}_{k_2}^{\{1,M\}}
\mathcal{D}_{k_2-k_1}^{\{1,M\}}
}~,
\end{eqnarray}
with $\mathcal{C}_{\textbf{5}} \equiv
\delta_{AB}{\rm Tr}(T^A_{\textbf{5}}T^B_{\textbf{5}})$, where $(T^A_{\textbf{R}}
T^A_{\textbf{R}})_{ij}$ is the quadratic Casimir operator for the generic irreducible representation ${\footnotesize{\textbf{R}}}$. 
For the quintuplet we find ${\rm Tr}(T^A_{\textbf{5}}T^B_{\textbf{5}}) = 10\delta_{AB}$.
We used the short-hand notation $\mathcal{D}_{k}^{\{n,M\}} \equiv (k^2 - M^2)^n$. 
Notice that the class of double-bubble corrections like the last diagram in the second row of fig.~\ref{fig:TwoLCanc} 
gives a vanishing contribution since  
it involves the trace in the isospin space ${\rm Tr}(T^A_\textbf{5}T^3_\textbf{5}T^B_\textbf{5}) =  0$. 
The diagram $\Pi_{S}^{(2)}$ vanishes, 
since the two integrations are factorized, and the integral  over $k_1$ does not possess any scale.
As far as the other diagrams are concerned, 
it is possible to show---retaining only corrections proportional to $M^2$---that $\Pi_{S}^{(1)} + \Pi_{F}^{(1)} + \Pi_{F}^{(2)} = - \Pi_{S}^{(3)}$, thus enforcing the cancellation.

The three-loop correction to the Higgs boson mass is 
\be
 \delta m_h^2 \sim 
 \frac{g^6}{(4 \pi)^6} M^2\, ,
\ee
with a coefficient in front which depends on many diagrams (see fig.~\ref{fig:HiggsCorrections}) and that we  take of O(1). The naturalness constraint on the mass $M$ is accordingly shifted by a factor $g/(4\pi)$ with respect to the two-loop result given by \eq{2loop} and becomes 15 TeV for the Majorana state and 42 TeV for the scalar. 

\vskip 1.3em
\textit{Relic density and DM mass.} 
MDM is a constrained model in which the only free parameter is the mass $M$ of the DM candidate. This is fixed by the thermal relic abundance---which is constrained by Planck data~\cite{Ade:2015xua}
 to  be
\be
\Omega_{\rm DM} h^2 = \frac{Y_{\infty} M s_0}{\rho_{c}^{0}h^{-2}}=    0.1188 \pm  0.0022\, , \label{relic}
\ee
where $s_0 \simeq 2.71\times 10^3$ cm$^{-3}$ and $\rho_{c}^{0}h^{-2} \simeq 1.05\times 10^{-5}$ GeV cm$^{-3}$
are the present entropy and critical energy density of the Universe while $Y_{\infty}$ is the asymptotical value of the DM comoving density.
In the usual freeze-out scenario, the value of $Y_{\infty}$ is controlled by a system of coupled Boltzmann equations describing, as the Universe gradually expands and cools,
 the evolution of the DM density driven by its interactions with SM particles.

\begin{figure}[!t!]
\centering
 \includegraphics[width = 0.45 \textwidth]{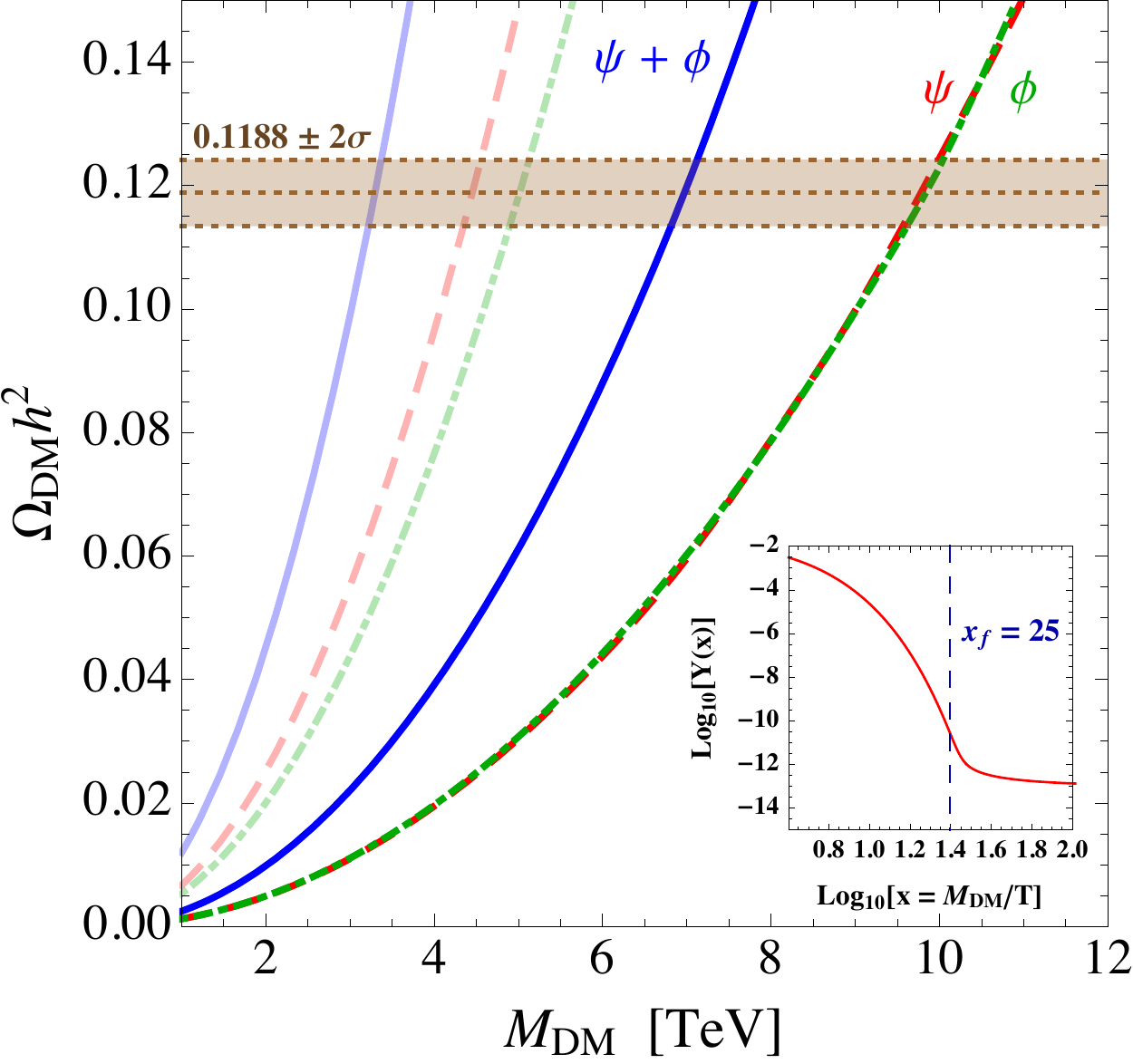}
\caption{ \textit{Thermal relic density of DM as a function of its mass $M$ for the two DM components $\phi$ and $\psi$  taken one at the time and together. Darker (lighter) curves with (without) the Sommerfeld enhancement. In the inset, the DM comoving density as a function of the variable $x=M_{DM}/T$  shows the freeze-out temperature $T_f$.
}}
\label{fig:RelicDensity}
\end{figure}

In our case we have two DM candidates, the scalar $\phi$ and the fermion $\psi$, with the same mass $M$, both $SU(2)_L$ multiplets with $n=5$; they thermally averaged  cross sections for co-annhilation  into SM states are given by, respectively~\cite{Cirelli:2005uq}:
\bea
 \langle \sigma v \rangle_\phi &=&  \frac{g^4}{64 \pi M^2 2 n} \left[ 3 - 4 n^2 + n^4 + \frac{n^2-1}{2} \right]  \\
 \langle \sigma v \rangle_\psi &=& \frac{g^4}{64 \pi M^2 2 n} \Big[ 2n^4 + 17 n - 19 \Big] \, . \label{x}
 \eea

The problem reduces to the solution of two separated Boltzmann equations for the two multiplets---since 
the only interactions between  $\phi^A$ and  $\psi^A$ 
are those mediated by gaugino exchange and the phase space of these interactions is kinematically closed if the mass splitting is neglected. 
For the values of masses we will find, the freeze-out temperature temperature $T_f \simeq M/25$ is above the EW phase transition and we can neglect  the mass correction
in \eq{split} induced by the Higgs vacuum expectation value.

By solving the Boltzmann equations (fig.~\ref{fig:RelicDensity}), we  find that  the relic density is saturated by the two DM states for a value $M\simeq 3$ TeV 
of their common mass (lighter lines, see caption).
 This result does not take into account the Sommerfeld enhancement of the cross section. 
 The Sommerfeld enhancement  is  a  non-relativistic  effect  generated  by  the  exchange  of  light  force  carriers 
 between the two annihilating DM particles~\cite{Hisano:2006nn}. 
 In the MDM scenario, the EW gauge bosons are the mediators responsible for this effect~\cite{Cirelli:2007xd,Cirelli:2015bda}.
 After including this correction---which we compute following closely~\cite{Cirelli:2015bda}---we conclude that the relic abundance~\eq{relic} 
 is actually saturated by a value $M\simeq 7 $ TeV (see fig.~\ref{fig:RelicDensity}). Such a value  is inside the naturalness limits found in the previous section.

\vskip 1.3em
\textit{Phenomenological signatures.} 
The model requires, in addition to the DM candidates, that gauginos and two higgsinos   be added to the SM states. As already discussed, these states combine to form a  neutralino $\tilde G^0$ and  charginos $\tilde g^+$ and $\tilde g^-$ with masses of the order of the EW breaking scale. The neutralino does not contain the hypercharged component and---because of the value of  its mass---it cannot be a DM candidate~\cite{ArkaniHamed:2006mb}. Searches for these states are under way at the LHC (see, for example, \cite{Khachatryan:2014mma}) as wino-like chargino and neutralino production within the minimal supersymmetric extension of the SM. 

The scalar components of the chiral multiplets $\Phi^a_{H_{u,d}}$ give rise to a two Higgs doublet model (2HDM), the phenomenology of which is actively under study at the LHC---primarily in the decay modes of the Higgs boson. Limits on the parameter space can be found in the literature (see, for instance, \cite{Chen:2013rba}). 

The model does not predict the existence of any colored super-partner.

\vskip 1.3em
\textit{Conclusions.} 
We have shown that a model with a distinctive phenomenology can be defined by requiring MDM to be natural. A partial implementation of SUSY  solves the problem. It stands as an example of a general approach to naturalness that addresses the little hierarchy problem which arises when new physics is added beyond the SM at an energy scale  around 10 TeV.

\vskip 1.3em
\textit{Acknowledgments.} MF thanks  SISSA  for the hospitality. We thank Marco Taoso for discussions and important remarks on the first version of the paper.



\end{document}